%% file: ms.tex
\documentclass[a4paper,11pt]{article}

\usepackage[plain]{fullpage}
\usepackage{graphicx}  %This enables the inclusion of pdf graphic files in figures
\usepackage{hyperref} % Make links in your document click-able NB: must be loaded before the caption-package
\usepackage{caption}
\usepackage{subcaption}
\usepackage{wrapfig} 
\usepackage{titling}
\usepackage{float}
\usepackage{listings}
\usepackage{geometry}
\usepackage{marginnote}
\usepackage[toc,page]{appendix}
\usepackage[utf8]{inputenc}
\usepackage[acronym]{glossaries}
\usepackage[style=alphabetic, backend=biber]{biblatex}
\makeglossaries

\title{ \acrfull{jwt} based client authentication\\
in \acrfull{mqtt}\\}
    
\author{\textbf Krishna Shingala\\
\texttt{krishna.shingala@nordicsemi.no}\\}
\date{August 2018}
\makeglossaries

\begin{document}

     \maketitle
     \begin{titlingpage}
        
        \vspace{5cm}
        \begin{abstract}
        This paper is an overview of \acrfull{jwt} and  \acrfull{tls} as two primary approaches for authentication of the things on the Internet. \acrfull{jwt} is used extensively today for authorization and authentication within the OAuth and the OpenId framework. Recently, the Google Cloud \acrshort{iot} has mandated the use of  \acrshort{jwt} for both HTTP and \acrfull{mqtt} protocol based clients connecting to the cloud service securely over \acrshort{tls}. \acrshort{mqtt} is the protocol of choice in \acrshort{iot} devices and is the primary focus of this paper as the application protocol. Another popular cloud platform \acrfull{aws} uses the \acrshort{tls} mutual authentication for client authentication. Any comparison provided here between the two approaches is primarily from a constrained device client perspective. 
        \end{abstract}
    \end{titlingpage}
    
    \newpage
    \section{Introduction} \label{intro}
    \input{01-Introduction}

    \section{JSON Web Token} \label{jwt}
    \input{02-JWT}

    \section{Message Queuing Telemetry Transport Protocol} \label{mqtt}
    \input{03-MQTT}
    
    \section{MQTT Client Authentication Schemes} \label{schemes}
    \input{04-Schemes}

    \section{Comparison of authentication schemes} \label{comparion}
    \input{05-Comparison}
    
    \section{Conclusion} \label{conclusion}
    \input{06-Conclusion}

    \newpage
    \appendix
    \section{Authentication with TLS} \label{tls_auth}
    \input{tls_authentication.tex}
    
    \newpage
    \input{ms.gls}
    \printglossaries
    
    \newpage
    \printbibliography
\end{document}

%% file: 01-Introduction.tex
The \acrfull{jwt}, defined by \cite{rfc-jwt} enable digitally secure representation and exchange of claims between two or more parties on the internet.

The \gls{jwt} have been used in the OAuth framework for authorization grants by the user of service to a third party. Such a grant enables third party applications to have access to users resources on the service. The OAuth framework is extensively used for web and mobile phone applications, and is specified in the \cite{rfc-oauth}. The use of \acrshort{jwt} within the OAuth framework is specified in the \cite{rfc-client-authentication}.

\begin{figure}[ht!]
\centering
\includegraphics[scale=0.48]{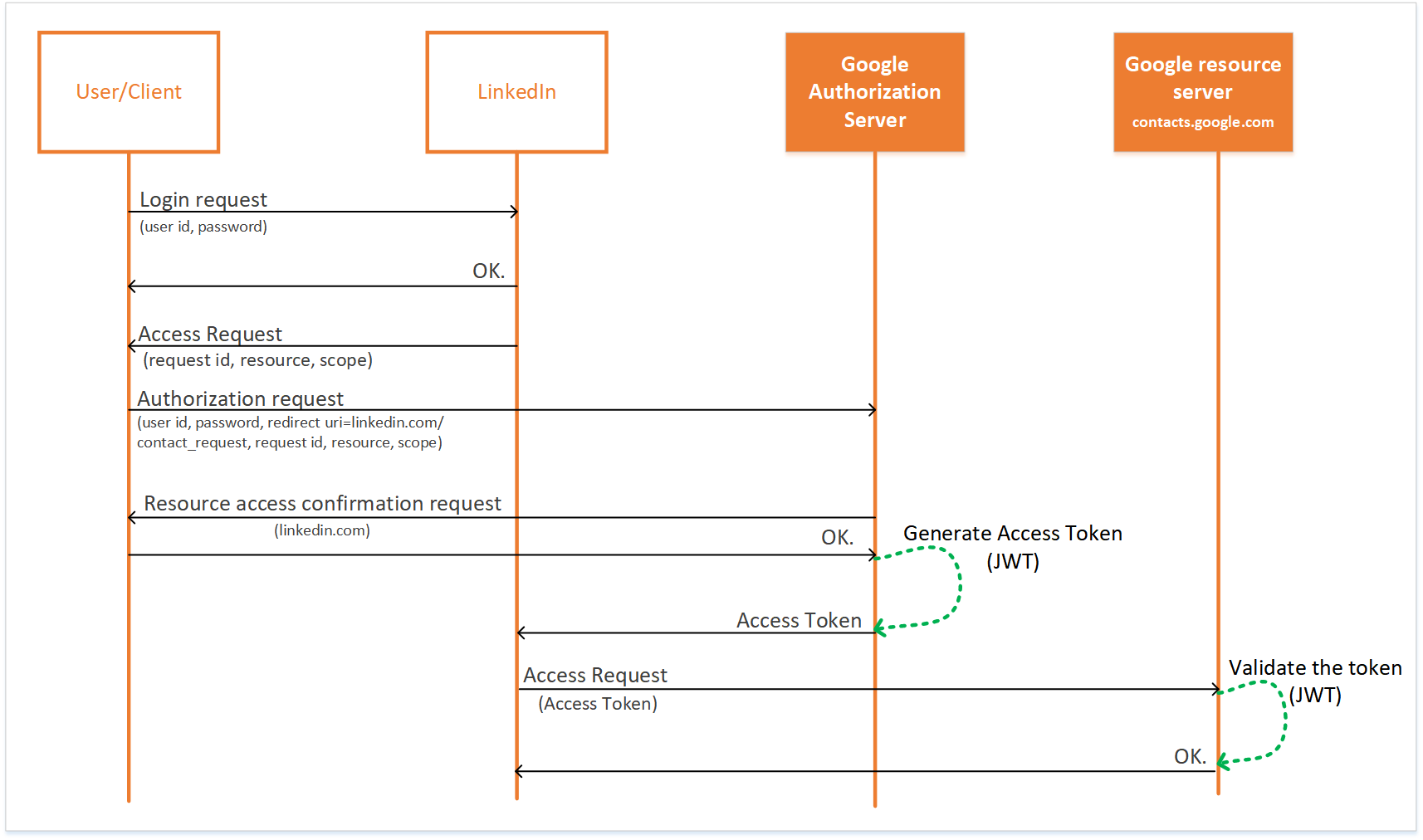}
\caption{Use of \acrshort{jwt} as access tokens in OAuth}
\label{fig:accesstoken}
\end{figure}

Figure \ref{fig:accesstoken} demonstrates an example and simplified usage of \acrshort{jwt} as access tokens used by a third party application to get user authorized access to resources of another service. Here LinkedIn is the third party application that requests access to users contacts (resources) of the user's Gail account.

The OpenID Connect extends the OAuth, and use of \acrshort{jwt} for authentication purposes. The OpenId Connect specification is available at \cite{openid-spec}.

The Google Cloud has now introduced use of \acrshort{jwt} for an \acrshort{iot} protocol \acrfull{mqtt} for authentication. The \acrshort{mqtt} specification is available at \cite{mqtt-spec}. The \acrfull{aws}, another popular cloud service employs the \acrshort{tls} Client Certificates as the primary mechanism for authenticating clients connecting to the \acrshort{iot} service. This paper discusses available mechanisms for authentication with \acrshort{mqtt}.

%% file: 02-JWT.tex
The \acrfull{jwt}, defined by \cite{rfc-jwt} enable digitally secure representation and exchange of claims between two or more parties on the internet. The claims are described in the \acrfull{json} format. The claims can then be encrypted, as \acrfull{jwe}, or, can be digitally signed or mac protected using the \acrfull{jws}. The \acrshort{jwe} specified in the \cite{rfc-jwe}. The \acrshort{jws} specified in the \cite{rfc-jws}. The \acrshort{json} format is specified in \cite{rfc-json}.

\begin{figure}[ht!]
\centering
\includegraphics[scale=0.7]{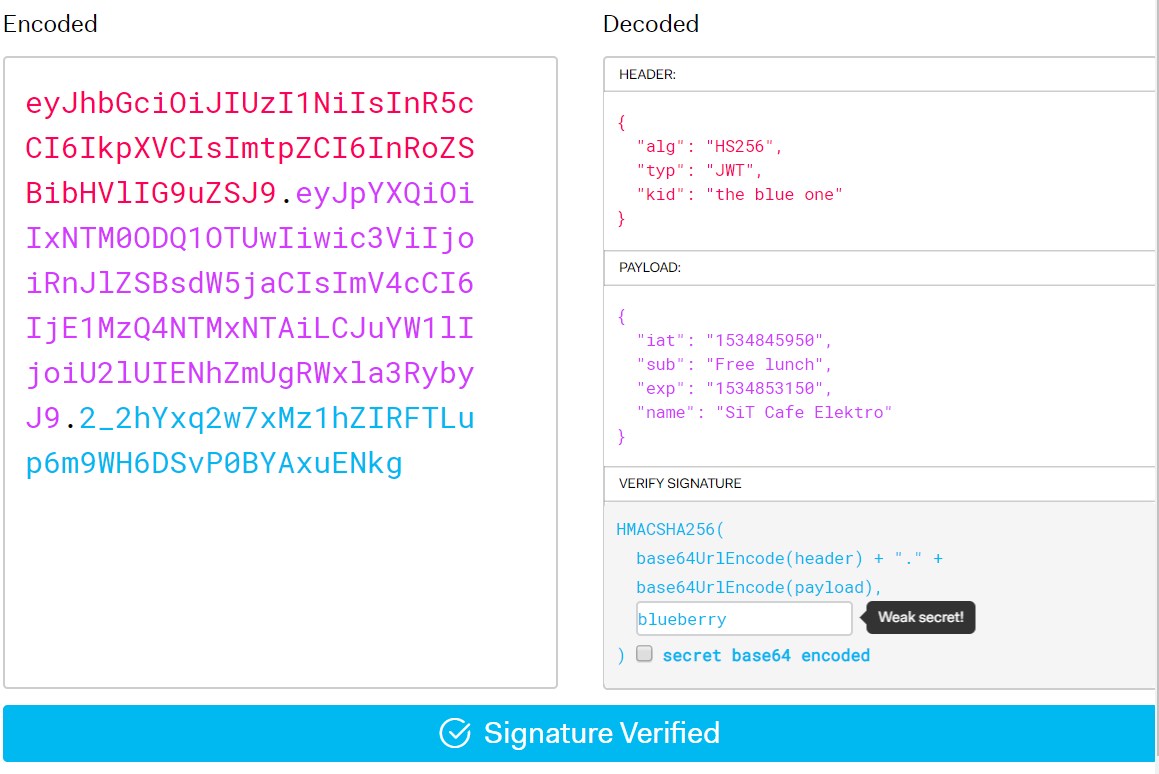}
\caption{Use of JWT for authorization grant in the OAuth framework}
\label{fig:jwt-sample}
\end{figure}

Figure \ref{fig:jwt-sample} demonstrates example uses of \acrshort{jwt}, \acrshort{jws} to be specific, for encoding a set of claims. The \acrshort{jws} consists of three parts:
\begin{itemize}
    \item the header, describes the primitives in \acrshort{json} format used for securing the claims.
    \item the payload, or the body, describes the claims in \acrshort{json} format.
    \item the signature or the message authentication code on the base64url encoded header and the payload.
\end{itemize}

Each field of the \acrshort{jwt} is base64url encoded and separated by a ".". The base64url is specified in the \cite{rfc-encoding}. The issuer of the claims, described by the "iss" field is the SiT Cafe Elektro and the subject of the claim, described by the "sub" field is free lunch. Further, the time of issue of the claim, is described by the "iat" (issued at) field. Similarly, the expiration of the claim is encoded in the "exp" field. The time format used to represent values of both "iat" and "exp" fields are as defined by the ISO-8601 standard. Here, the free lunch claim issued at Tuesday, August 21, 2018 10:05:50 AM UTC time and expires at Tuesday, August 21, 2018 12:05:50 PM UTC time. These claims are message authenticated using HMAC using SHA-256(HS256) under the shared secret "blueberry". 

The HS256 scheme, uses the same key for signing and verification and is a message authentication scheme. Signature schemes, based on public key cryptography schemes use a signing key (private to the signer) to sign the messages and a verification key to verify the signatures, and hence authenticate the peer. \acrshort{rsa} and \acrshort{ecdsa} schemes are supported in \acrshort{jwt}. The algorithms used for securing the \acrshort{jwt} have been defined in the \cite{rfc-alg}.  In later sections, when referring to \acrshort{jwt} based schemes is referred to, public key based signature schemes are implied.

Clearly, the \acrshort{jwt} server very useful for issue of digital tokens that, if valid, can be exchanged for access to services. 

It is important to note that \acrshort{jws} mus be exchanged on a secure channel to avoid being stolen and misused by sniffing party. \acrfull{tls} is commonly used for establishing a secure channel for exchange of the \acrshort{jws} as access tokens.

%% file: 03-MQTT.tex
\begin{figure}[ht!]
\centering
\includegraphics[scale=0.5]{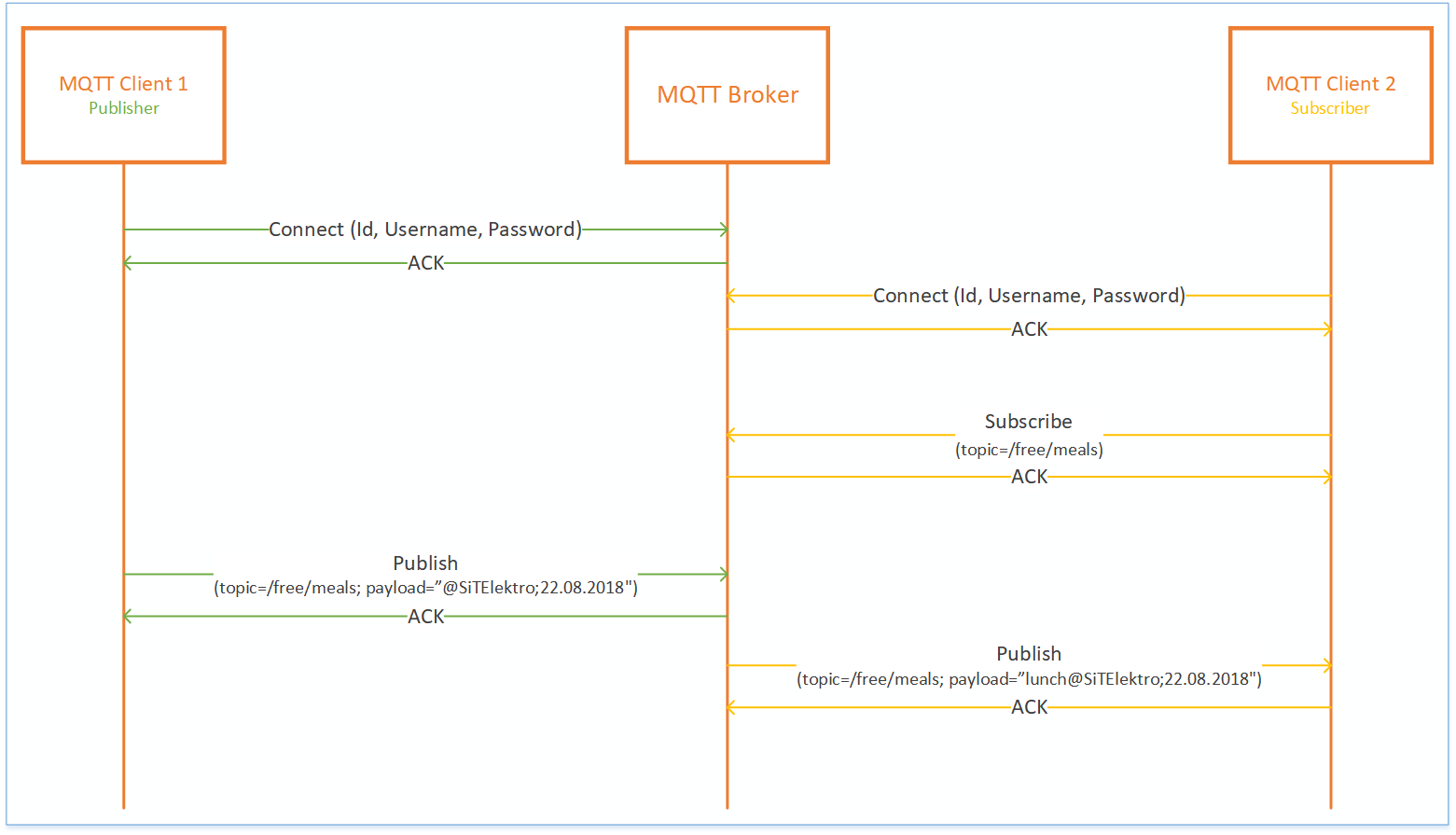}
\caption{MQTT Protocol Overview}
\label{fig:mqtt}
\end{figure}

The \acrshort{mqtt}, an \acrfull{oasis} standard, is a lightweight protocol for machine to machine communication. All machines, referred to as the client communicate through a central server referred to as the broker. The publish-subscribe pattern is used for message exchange between the broker and the client. The clients can be publishers, subscriber or both. All clients must identify themselves uniquely when connecting to the broker. Figure \ref{fig:mqtt} depicts two clients connecting to the broker. One of the client is the data source and publishes data, while another is a subscribes and subscribes to messages published message on a specific topic(s).

The \cite{mqtt-spec} defines many concepts and mechanisms to enable detection of inactive clients, publishing the last will and testament of such clients, graceful disconnection etc. However, only the concepts necessary relevant for the discussion on client authentication are described here.

\subsection{Transport}
The \acrshort{mqtt} is defined over the \acrshort{tcp} transport. \acrshort{tcp} port 1883 is reserved for the \acrshort{mqtt} protocol. In case \acrshort{tls} is used for securing communication between the client and the broker, then the \acrshort{tcp} port 8883 is used. 

A new specification,\acrfull{mqtt-sn}, adapted for the \acrshort{udp} transport has been defined. See \cite{mqtt-sn-spec} for details. This version is not discussed in this paper. 

\subsection{Publish and Subscribe}
A client can publish data to a broker using the \acrshort{mqtt} publish message. A broker can similarly send data to a client via a publish message. Each publish message contains a topic field that identifies the data being published, for example, whether the data is a temperature measurement or a \acrshort{gps} may be segregated based on the topic. Similarly, if the client is interested in receiving certain measurements, then it can subscribe to its topic(s) of interest. The topic and payload are of variable length.

The specification does not mandate any topics nor format of the topics. However, it does define wild card characters that permissible when subscribing to topics. The specification also does not mandate any payload formats that shall be supported by the clients.

Therefore, the specification leaves much to the implementation, and the clients must comply with the broker that it wishes to communicate to.

It is worth noting that two client never talk to each other directly.

\subsection{Client Authentication and Authorization} \label{authentication and authorization}
The \cite{mqtt-spec} defines optional authentication of the clients connecting to the broker using user name and password of the connection request. The primitives used for the user name and password is an implementation choice left to the entity that deploys the \acrshort{mqtt} based service. 
Based, on the configured access control policy on the broker, clients may have authorization to publish and/or subscribe to only certain topics. 
The specification does mention authentication of the clients using the \acrshort{tls} mutual authentication scheme. Therefore, it is not necessary that all authentication schemes must use the user name and password field of the connect request.

The choice of authentication and authorized access is left to the implementation and not mandated by the specification. Some implementations of \acrshort{mqtt} make issues with leaving much to the implementation clear. These implementations, by default, allow unauthenticated clients to publish data to authenticated subscribing clients that authenticate the server using \acrshort{tls}. Therefore, the authentication of the server provides little value to the subscribers.

\subsection{Keep Alive}\label{keepalive}
The specification defines a periodic keep alive message be sent by the client to the broker. This is intended for the broker to detect any clients that are no longer reachable due to battery failure, loss of network or other reasons. 

%% file: 04-Schemes.tex
The specification of \acrshort{mqtt},  allows for implementation specific client authentication scheme.  Section \ref{authentication and authorization} already summarizes client authentication and authorization as defined in the \acrshort{mqtt} specification, \cite{mqtt-spec}. The specification focuses mostly on client authentication and not broker (server) authentication. However, server authentication is assumed achievable using \acrshort{tls} server authentication schemes.   

Before, we discuss the client authentication schemes, it is recommended to familiarize with  \acrshort{tls}. Knowledge of certificate authentication of the \acrshort{tls} server and the \acrshort{tls} client will be particularly useful to appreciating how some of the schemes work, and to compare them with the others. The version 1.2 specification of \acrshort{tls} is available at \cite{rfc-tls-1.2}. \cite{kap} provides a good overview of \acrshort{tls} and an excellent summary of attacks on \acrshort{tls}. \ref{tls_auth} servers to provide a quick refresh of server and client authentication during \acrshort{tls} handshake.

%\begin{figure}[h!]

Since, the specification leaves much to implementation and choice, all possibilities are not considered and compared. This research is limited to comparing the schemes listed below:
\begin{itemize}
    \item User name and password
    \item Client Authentication using \acrshort{tls}
    \item Client Authentication using \acrshort{jwt}
\end{itemize}

On note for the subsequent sections is that the service (the cloud) may consist of many components. Components like the \acrshort{mqtt} broker to communicate with the sensor devices that implement\acrshort{mqtt}clients, a user data base or an \acrfull{iam} system for maintaining the users identity, authentication and authorization information, a portal to register the users, a portal to monitor the users data from the clients and/or the state of the clients etc. The service therefore is a composite of various components. However, in this paper, only two components are considered, the \acrshort{mqtt} broker and the user data base or the \acrshort{iam}. Note that for the client, the broker symbolizes the service. 

\subsection{User name and password}\label{username_password}

The \acrshort{mqtt} specification \cite{mqtt-spec} provisions for user name and password fields in the connect request message from the client. Both the fields are optional. And presence of each field is individually indicated in the flags field of the message. See section 3.1 of \cite{mqtt-spec} for details.

\begin{figure}[ht!]
\centering
\includegraphics[scale=0.6]{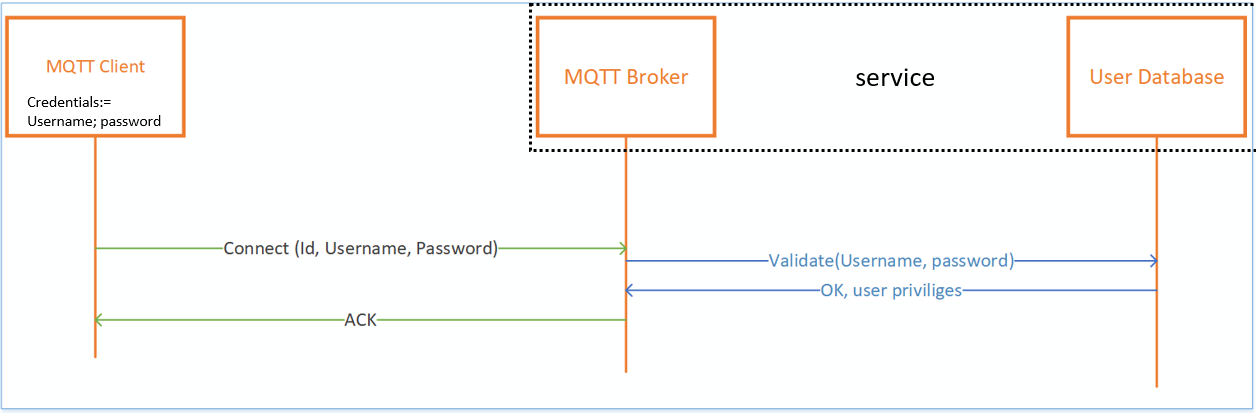}
\caption{MQTT Client Authentication: User name and password}
\label{fig:mqtt-username-password}
\end{figure}

The figure \ref{fig:mqtt-username-password} provides a simplified view of a client connecting to \acrshort{mqtt} based service, that uses the user name and password field for authentication. Authorization is derived from the authentication of the user name. 

\subsection{Prerequisite}
For this authentication scheme, following prerequisites must be fulfilled.
\begin{itemize}
    \item User name and password are registered with the service.
    \item The user name and password have been provisioned in the client.
\end{itemize}

This method is simplistic, and requires the client to share its identity and secret to the broker. And that once the credentials are compromised, any body could start impersonating the client. If this scheme must really be used, then, at least two precautions must be taken to:
\begin{itemize}
    \item Encrypt the communication between the client and the broker.
    \item Authenticate the server before sending any credentials to it.
\end{itemize}

Both of these are easily addressed by use of TLS with server authentication scheme.

Note that this method may never be deployed commercially, however, serves as a good basis for understanding how use of resources on the brokers can be authorized by a simple use of user name.

It should be noted that the specification uses the client id and the user name as distinct concepts without much elaboration on intended use. One of the interpretations could be that a single user can log in from multiple devices. With this interpretation one could argue if the authentication is really the user authentication or client authentication.

However, it is mandated that the client id be unique across clients, and the password field could be supplied even without the user name field. This leaves possibility of individual client's authentication open. It seems intentional to permit both models in the specification. However, this is subjective to the readers understanding.

\subsection{Client Authentication using TLS} \label{mqtt_tls_client_auth}

\begin{figure}[ht!]
\centering
\includegraphics[scale=0.6]{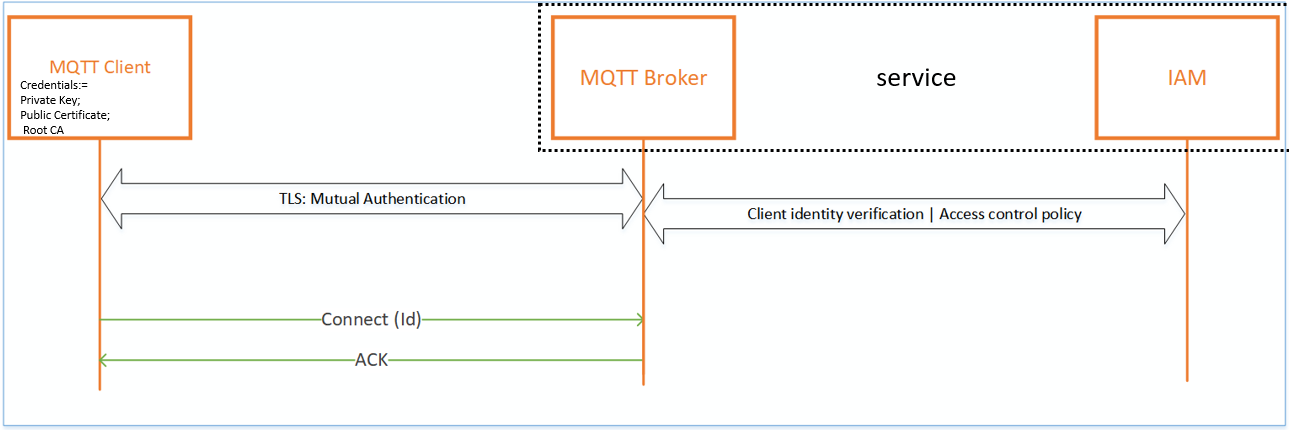}
\caption{MQTT Client Authentication: TLS Client Authentication}
\label{fig:mqtt-tls-client-authentication}
\end{figure}

The \acrshort{mqtt} broker relies on the \acrshort{tls} client certificates to establish the identity of the client, and authenticate it. The access control policies are applied based on the identity of the client, the public certificate. The figure \ref{fig:mqtt-tls-client-authentication} illustrates the authentication of a \acrshort{mqtt} client authenticated using \acrshort{tls} before the \acrshort{mqtt} connection. This method is not unique to \acrshort{mqtt}, and \acrshort{tls} based client authentication is available to any protocol that uses \acrshort{tls} for establishing a secure channel. This method is already popularly used with the \acrshort{http} based applications.

\subsection{Prerequisite}
For this authentication scheme, following prerequisites must be fulfilled.
\begin{itemize}
    \item The client is provisioned with its private key and public certificate. In addition, root \acrshort{ca} certificate is provisioned to authenticate the server.
    \item Client's public certificate is registered with a service and necessary access control policies attached.
\end{itemize}

The \acrfull{aws} cloud service deploys this method as the default and recommended mechanism to authenticate the both \acrshort{http} and the \acrshort{mqtt} based clients. The client certificates registered with the \acrshort{aws} are signed by a root \acrshort{ca} that \acrshort{aws} trusts. In fact, \acrshort{aws} implements its root \acrshort{ca} based on the geographic region. The certificate chain used by \acrshort{aws} is not very long.The documentation of integrating a device to with the \acrfull{aws} IoT is available at \cite{aws-iot}. Protocol specific documentation is available at \cite{aws-protocols}. Additional documentation on security and identity is available at \cite{aws-security}. 

It is important to note that \acrshort{aws} employs many components, each component delegated a specific task. And achieving communication with the \acrshort{mqtt} broker has the prerequisite of setting up other components up correctly. An example of such a component is configuration of the \acrshort{iam} component. Documentation is available at \cite{aws-iam}. The \acrshort{aws} \acrshort{iam} provides a verify fine grain control of access control policies per certificate. 

This method, enjoys the merit of being a well established standard available to application protocols. And therefore, could be more studied and attacked as against the others schemes.

One observation from experience of implementing \acrshort{tls} mutual authentication on Cortex-M based embedded device is that the RAM and CPU requirements when implementing \acrshort{tls} peak the handshake are quite high. In fact \acrshort{tls} extension max fragment length had to be enabled in order to reduce the RAM requirements for the input and the out record sizes. Hence, this method can be resource intensive and impractical for small embedded devices. This could be a reason why \acrshort{aws} may have opened up for other possibilities for authentication of the \acrshort{iot} devices.

\subsection{Client Authentication using JWT}\label{jwt-based-auth}
\begin{figure}[ht!]
\centering
\includegraphics[scale=0.6]{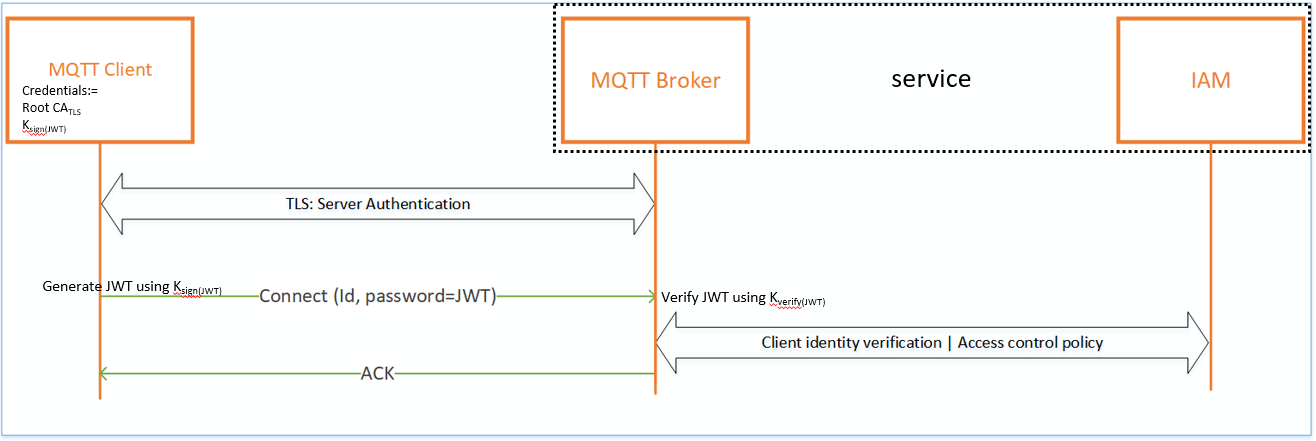}
\caption{MQTT Client Authentication: JWT}
\label{fig:mqtt-jwt}
\end{figure}

The \acrshort{mqtt} client sends the broker a \acrshort{jwt} in the password field of the connect message to the broker. This is the first message that the client send to the broker. The \acrshort{jwt} contained in the password field contains the claims like the time of issue, expiration date and any other claim defined by the service and a signature of these claims, any header fields. The signature servers as a proof of possession of the signing key or the private key. The signature is verified using the verification key or the public key that is already registered with the service. Therefore, the client's identity is its public certificate. However, this identity is never sent out during the communication. The \acrshort{jwt} may include a hint on the key to be used for verification. An illustration of this method is depicted in the figure \ref{fig:mqtt-jwt}.

\label{mqtt_jwt_client_auth}

\subsection{Prerequisite}
For this authentication scheme, following prerequisites must be fulfilled.
\begin{itemize}
    \item The client is provisioned with it a signing key (the private key). In addition, root \acrshort{ca} certificate is provisioned to authenticate the server.
    \item Client's public certificate is registered with a service and necessary access control policies attached.
\end{itemize}

This mechanism of authenticating the clients is used by the Google Cloud IoT Core service. The service mandates inclusion of the audience claim identifying the project in the \acrshort{jwt}. The mandated header fields include the algorithm and type.  The signature schemes mandated to be supported for \acrshort{jwt} are RS256 and ES256. The service does not mandate the verification keys provisioned on the server for \acrshort{jwt} to be signed by any \acrshort{ca}. These can be self signed. The Google cloud mandates refreshing the \acrshort{jwt} at least once every hour. It is not specified if any user or project specific claims can be included in the \acrshort{jwt}. Full documentation on how to integrate a device with the Google Cloud IoT using \acrfull{mqtt} is available at \cite{google-mqtt}. The documentation on use \acrshort{jwt} for authentication is available at \cite{google-jwt}.

Use of \acrshort{tls} is a must as otherwise, the \acrshort{jwt} stand the risk of being stolen and misused.
The Google Cloud service mandates use of \acrshort{tls}. The server is authenticated by the server certificate and also confidentiality of communication between the broker and the client is ensured. One observation is that the service uses a long \acrshort{ca} chain. This has implications on bandwidth, RAM and time taken to verify the server on the client side, which may be a small embedded device.

The client id format are mandated by the service. This is unlike the \acrshort{aws} service that allows the developers to define the length.

%% file: 05-Comparison.tex
This section compares the authentication schemes described in \ref{schemes}. Not all attributes used for comparison are security properties or goals. However, these are included for as they integrate the application layer protocol, and implementation details and challenges with the schemes. 

It is important to note that no formal security analysis has been performed for comparing the schemes. Tamarin-prover available at \cite{tamarin-prover} was investigated for the purposes of analyzing use of \acrshort{tls} and \acrshort{jwt} for \acrshort{mqtt} together. However, the investigation never materialized mostly because of time constraints. Also, there was no straight forward way of combining protocols and primitives for analysis exist.

\subsection{Confidentiality}
Confidentiality here is considered not as a security goal, but rather whether it is a prerequisite for authentication. The \acrshort{jwt} based authentication described in the section and \ref{username_password} \ref{jwt-based-auth} must be performed on a confidential, and server authenticated channel, to be of any value. Else, the token or the credentials may be stolen and used by anybody to get unauthorized access to the resources on the server. 

The tokens are stateless, and do no include any client and/or protocol information. Of course the access to resources are time limited, but, provide enough opportunity to exploit the service by means of impersonating a legitimate client of the service.

The \acrshort{tls} client authentication scheme has no such requirement. However, as a consequence, compromise the client's privacy as a consequence. See \ref{privacy} for details.

\subsection{Server authentication}
Again, this discussed more as a prerequisite rather than a goal for all the schemes. Providing credentials to an unauthenticated server compromises not just the client, but could end up having wider implications on the system. If a client provides its credentials to an adversary, then, the adversary can impersonate the client publish messages that can cause harm through the devices that subscribe to these published messages. Note that the subscribers never authenticate the source of the published messages, but rather rely on the server to take necessary measures for entity and message authenticity.

\subsection{Client Privacy}\label{privacy}
Use of \acrshort{tls} client certificates for authentication described in \ref{mqtt_tls_client_auth} implies that the client's identity, its public certificate is sent as plain text. The \acrshort{tls} handshake is not complete yet, and hence the keys needed for bulk encryption are no established yet. A method to overcome this has been described in section of \cite{kap}. The method proposed is to first establish a confidential link with the server, and then renegotiate the session. Renegotiation procedure is apart of the \acrshort{tls}, however, must be supported by both the server and the client to successfully use this feature. Attacks that target this renegotiation procedure are also detailed in \cite{kap}.

The use of \acrshort{jwt} for authentication described in \ref{mqtt_jwt_client_auth} fairs better in this aspect, as the the \acrshort{jwt} is sent over \acrshort{tls} encrypted channel. Also, the client's identity  is never directly transferred to the server. Rather the client identity and authentication is implied by a valid signature on the token.

\subsection{Credential/Key management}
One of the major challenges with deployment of any security scheme is the key management.

In \acrshort{tls} based authentication scheme, each device, ideally, should be provisioned with unique key pair - the private key, and the corresponding public certificate chain to authenticate itself, and the root \acrshort{ca} to be able to authenticate the server.

In \acrshort{jwt} base authentication scheme, each device, ideally, should be provisioned with he private key for signing the tokens and the root \acrshort{ca} to be able to authenticate the server during the \acrshort{tls} handshake.

For each scheme, it is recommended to have more than one key pair to be able to revoke them in case of compromise. 

Provisioning of private keys must be performed in a secure environment and managing them is an existing challenge that still requires o be addressed. Further, checking the revocation status of certificates may not be implemented in embedded devices. This choice may then lead to continued communication of devices with a compromised server. Millions of devices can hence be compromised due to a compromised server and impacts of such large scale compromises are hitherto unknown as the \acrshort{iot} is still reach its full potential. Updating the system of remote sensors can definitely have a cost impact on the businesses.

Strong recommendations, clear guidelines and swift and transparent measures in case of compromise by service providers may already be some of the steps that could be taken by the service providers.

\subsection{Requirement for Time Synchronization}
Use of \acrshort{jwt} based authentication requires implementing a time service. Use of time adds the needed randomness in \acrshort{jwt} tokens. Else, the signature scheme being deterministic would always have the same signature. Further, use of time ensures freshness of the authentication token. 

\acrshort{tls} based scheme does not have any such requirements as nonce provided by server and client are used during the handshake.

Connection is rejected by the Google cloud if the "issued at time" did not match its expectation. This means that there can be denial of service attacks launched by impersonating a time server. Typically an \acrfull{ntp} server is contacted by an embedded device to get absolute time and typically, only a subset referred to as the \acrshort{sntp} is implemented on these constrained devices. The specifications of \acrshort{ntp} and \acrshort{sntp} are available at \cite{rfc-ntp} and \cite{rfc-sntp} respectively. The \cite{rfc-sntp} describes on security measures being too elaborate and/or complex to be used in simpler devices - this being very true for small embedded devices. Use of \acrshort{jwt} based authentication is hence creating dependency on another network based service that has its own vulnerabilities.

\subsection{Session interruption}
The mandate by the Google cloud service to refresh the \acrshort{jwt} token periodically. To fulfill this requirement, the client must be disconnected and connected back with a fresh token as there are no existing mechanisms to refresh the security token. This, therefore translates into the requirement of reestablishing the \acrshort{tls} session. Such an interruption may be undesirable in some use cases. Undesirable, due to latency introduced, or the cost of creating a session or both.

While refreshing the token seems like a good idea when used as cookies in the browser or in case authorization grants enable by OAuth. It is unclear what the objective is refreshing the token when used for authentication of \acrshort{mqtt} clients is. And therefore, this additional cost and disruption becomes more undesirable.

If the security objective is established, then perhaps a scheme to enable refreshing of tokens with interrupting the sessions could be useful.

\subsection{Cost on the client}
The client devices that comply with the authentication schemes dictated by the cloud service, are many times constrained in terms resources (RAM, bandwidth, flash and computational capacity and available power), and may further requirements on latency on sending measurements or reacting to commands from the cloud (turn on light bulb) for example.

A study of implementation cost of the various both the \acrshort{tls} and \acrshort{jwt} based schemes is needed and unavailable as of today.

Choice of deployment schemes like the length of \acrshort{ca} used for the server certificates has a direct impact on the constrained client. Google cloud service uses a very long ca chain, that must be first received and stored by the wireless sensor and then validate. The \acrshort{aws} uses only one \acrshort{ca} for the servers.

Further, certain features like he resumption of \acrshort{tls} session may benefit the constrained devices, however, are not always supported by the cloud service due to challenges with sharing the session tickets across load balanced servers. An investigation of how this could be better enabled in the servers, and hence be used by clients may be useful.

\subsection{Access Control}
It is important to notice the subtle difference in when the \acrshort{iam} is contacted in the \ref{fig:mqtt-tls-client-authentication} and the \ref{fig:mqtt-jwt}. In \acrshort{jwt} based scheme, since th client authentication is not established until the \acrshort{mqtt} connect request arrives from the client, the default access policy must be configured correctly to ensure no opportunities for unauthorized access open up between the \acrshort{tls} handshake completion and \acrshort{mqtt} connection.

Further, in this scheme, it is unclear how clients may be authorized to allow only certain protocols. Note that the protocols are typically are deployed in distinct port numbers. Hence, for \acrshort{tls} based scheme, based on protocol port and the client certificate, it may be possible to assert even before the handshake completes if the client should be allowed access to the service on the port.

\subsection{Known plain text attacks on TLS}\label{known plain text attacks}

The \acrshort{mqtt} by design is lightweight with limited set of messages. Each message has fixed header, may have variable header and may contain variable length payload.  

Also, the order of the messages can be determined, the first message shall always be the connect message. There is a periodic keep alive message etc.

The connect message fixed header and variable parts by specification, but the variable parts are fixed for a particular client. The keep alive message has no variable components and is sent out periodically.

Therefore, knowing the structure of the message, and most parts of the message, and their use in the protocol could be exploited for chosen plain text attacks on the ciphers used for the \acrshort{tls} connection with the broker.

Note that use of \acrshort{jwt} in the initial connect message helps only in introducing certain variation to the initial message. And this too, if an only if, the \acrshort{jwt} itself introduced a variable, like the time of issue or expiration and/ or other random component.

Therefore, a method to analyze the \acrshort{tls} with \acrshort{mqtt} may be useful in determining how secure the connection between the client and the broker really is.

\subsection{End to end security}

Most cloud providers including the \acrshort{aws} and the Google cloud claim end-to-end security. In fact many practitioners have the notion that use of \acrshort{tls} implies that end-to-end security is achieved. And this may be right. 

However, consider the use case where an \acrshort{mqtt} subscriber either turns on/off the light based on luminosity or illumination detected by the publishing clients. The subscribing client has no way of ensuring that each of the published message was in fact valid. The validity here may be in sense of time, and or in sense of who is allowed to publish illuminations. The trust is placed entirely on the server. This could, by design of \acrshort{mqtt} is the single point of failure. 

Possibilities of using \acrshort{jwt} in every publish message to asset the source of the message, that the subscriber can verify independent of the server before processing the request could be interesting to research.

Formally defining these security goals are recommended prerequisite to and analyze solutions.

%% file: 06-Conclusion.tex
\acrshort{jwt} brings about a new method of authenticating the client that can be easily integrated in the existing implementations of the \acrshort{mqtt}. However, this new mechanism does not simplify any of the existing challenges of key management. Both \acrshort{tls} and \acrshort{mqtt} require use of public and private key pair to be provisioned into the devices. Google infact recommends use of key rotation and can support up to 3 keys per device. And, since both schemes rely on the root of trust via the \acrshort{ca} chain for server authentication, the two schemes inherit the risks involved with use of certificate chains. The \acrshort{jwt} based schemes does protect client privacy, and this is a clear advantage over the \acrshort{tls} scheme. However, this advantage however be  leveled with widespread support for \acrshort{tls} 1.3.

Cost of implementing \acrshort{jwt} in constrained devices is not known entirely either. A study of cost weighed against the benefits may be important to help make choices in products.

All schemes rely on secure channel established using the \acrshort{tls}. \acrshort{tls} being a popular and widely used security protocol on the internet, is most researched and also may be most attacked. A compromise of \acrshort{tls} may impact either of the schemes. It is therefore important to ensure that practitioners choose strong ciphers. It is also critical evaluate and test the \acrshort{tls} library used in implementations. Also, building mechanisms to ensure patching of any vulnerabilities found in implementations quickly may help mitigate some of the risks in deployed products.

Also, enabling study of security protocols and primitives in an application protocol may be useful, and needed to expose any vulnerabilities exposed by the choice of primitives or the nature of the application protocol. As mentioned in \ref{known plain text attacks}, \acrshort{mqtt} being predictable in payload sent from the client to the server may enable known plaint text attacks on \acrshort{tls}. A good models and tool enabling such analysis may be worth research to install faith in security measures deployed in \acrshort{iot} systems. Also, many times when analyzing protocols, it is assumed that a fresh time stamp is just available in the system. This may not be true for embedded devices. And an attack on time service of the device may open up for new opportunities of attacks on the device. The most obvious category being the denial of service attack.

More \acrshort{jwt} based authentication schemes have been proposed for \acrshort{iot} devices. These schemes are not limited or bound to use of any particular application protocol and are targeted for smart home applications. For example, \cite{personal-oauth} suggesting setting up a personal OAuth servers on mobile phones that can issue tokens for the users things connecting to the cloud. Another paper, \cite{imei-based-authenitcation}, suggests use of the home gateway device to issue security tokens to other devices to get needed authorized access to services. Both these solution, may simplify the use key management problem for constrained devices. These proposals may have not scale to use cases outside of smart home and, may also be putting to much trust on devices like Gateways and mobile phones which themselves are subject to being compromised easily. However, these proposals do show that there are many possibilities to authenticating \acrshort{iot} devices and many varied objectives, so one solution may not fit all use cases.

%% file: tls_authentication.tex
\begin{figure}[h!]
\centering
\includegraphics{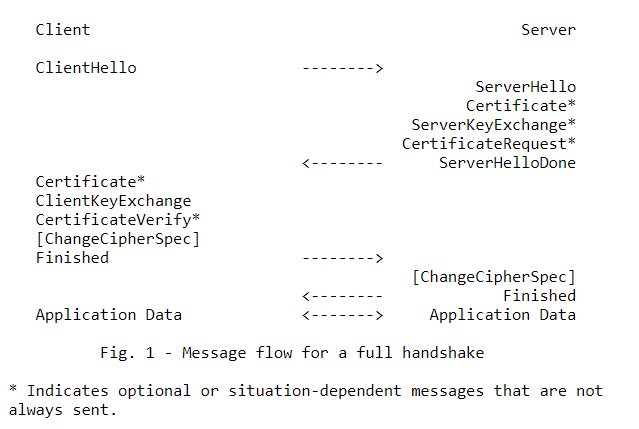}
\caption{TLS Handshake\\\texttt{Source: \cite{rfc-tls-1.0}}}
\label{fig:tls}
\end{figure}

The figure \ref{fig:tls} provides an overview of the handshake protocol of the \acrshort{tls}. The handshake protocol is used for authentication and key establishment. 

For server authentication, the client connecting to the server is provisioned with a root \acrshort{ca} certificate. During the handshake, the server then sends its certificate along with the certificate chain in the Certificate message that follows the ServerHello message. The client can now verifies the signature and validity period of each of the certificates in the certificate chain. The last certificate in the chain must either be the root \acrshort{ca} that the client is provisioned or signed by the this root \acrshort{ca}.

The client cannot send its own certificate unless the server requests it using the "CertificateRequest" message. When requested, the client, sends its certificate to the server in the "Certificate" message that follows the "ServerHelloDone" in the \ref{fig:tls}. The client is authenticated by the "CertificateVerify" message that contains signature on hash of all the handshake messages until this message.

Note that, the client cannot send its certificate if not requested by the server and, cannot demand that the server sends it certificate. It can however, terminate the handshake using the alert protocol.